\documentclass[prd,aps,showpacs,preprintnumbers,amssymb]{revtex4}
\usepackage{axodraw}
\usepackage{color}
\usepackage{epsf}

\def\e3p{$\eta \rightarrow 3 \pi$}

\begin{document}
\title{%
\hfill{\normalsize\vbox{%
\hbox{}
 }}\\
{  A solution to the naturalness problem}}

\author{Amir H. Fariborz
$^{\it \bf a}$~\footnote[1]{Email:
 fariboa@sunyit.edu}}
\author{Renata Jora
$^{\it \bf b}$~\footnote[2]{Email:
 rjora@theory.nipne.ro}}

\affiliation{$^{\bf \it a}$ Department of Mathematics/Physics, State University of New York, Polytechnic Institute,  Utica, NY 13504-3050, USA}
\affiliation{$^{\bf \it b}$ National Institute of Physics and Nuclear Engineering PO Box MG-6, Bucharest-Magurele, Romania}

\date{\today}

\begin{abstract}
We present a solution to the naturalness problem of the electroweak scale based on a change of variable in the Fourier space  of non supersymmetric nature that transforms a boson into a fermion and viceversa. This is exemplified for that part of the standard model Lagrangian that contains a Dirac fermion coupled with the Higgs field. The full Lagrangian which contains both the initial particles  and the partners obtained through this method is invariant under the symmetry associated to this change of variables and is free of quadratic divergences to the scalar particles present in the Lagrangian. The partners appear mostly as "off-shell" states thus explaining their experimental absence at any particle accelerators.

\end{abstract}
\maketitle

\section{Introduction}

Naturalness of the electroweak sector of the standard model \cite{Weinberg}, \cite{Susskind} ,\cite{Hooft} ( for reviews see \cite{Feng}, \cite{Dine}) has been at the forefront of theoretical research for many years. The discovery of the Higgs boson particle by the remarkable
Atlas \cite{Atlas} and CMS \cite{CMS} experiments together with the absence of any experimental sign of beyond the standard model physics  has put some strains on the most widely accepted  models that might explain naturalness. Among these we just mention technicolor models \cite{Weinberg}-\cite{Susskind}, little Higgs \cite{Kaplan}, \cite{Georgi}, \cite{Hamed},  Lee-Wick theories \cite{Grinstein}, large extra dimensions \cite{Hamed1}, \cite{Hamed2},
warped extra dimensions \cite{Randall1}, \cite{Randall2} and the most  acclaimed of all, the supersymmetric extensions of the standard model \cite{Maiani}-\cite{Seiberg}. The lack of superpartners at the LHC has lead to some alternatives to the standard lore.

The naturalness problem of the electroweak sector of the standard model is given by the magnitude of the quadratic divergences to the Higgs boson mass as compared to the mass itself. For a theory with multiple scalars this would translate into quadratic divergences to any scalar in the theory.  In supersymmetry one associates to each particle of the standard model a superpartner of different spin but with the same other quantum numbers. It is the mere presence of the superpartner and of the supersymmetric coupling that solves the naturalness problem because the superpartners introduce quadratic divergences with the same magnitude and opposite signs. However supersymmetry needs to be broken and the masses of the superpartners must be large as no experiment ever detected any trace of them.

One possible solution to the naturalness problem  that we shall adopt here  is to find versions of the standard model that have some of the features of the supersymmetric models but where the "superpartners" do not appear in most interactions as on-shell states.   We  regard this as a first attempt to build an extended standard model free of quadratic divergences and where all other particles that are introduced beside those of the standard model are present as "off-shell" fluctuations" but not as final states in the interactions.

The starting point of our proposal is the supersymmetric Wess Zumino Lagrangian,
\begin{eqnarray}
{\cal L}=-\partial^{\mu}\Phi^*\partial_{\mu}\Phi+i\bar{\Psi}\bar{\sigma}^{\mu}\partial_{\mu}\Psi,
\label{wess324}
\end{eqnarray}
which is invariant under the supersymmetry transformation:
\begin{eqnarray}
&&\delta \Phi=\epsilon\Psi
\nonumber\\
&&\delta\Phi^*=\epsilon^{\dagger}\Psi^{\dagger}
\nonumber\\
&&\delta \Psi=-i\sigma^{\mu}\epsilon^{\dagger}\partial_{\mu}\Phi
\nonumber\\
&&\delta\Psi^{\dagger}=i\epsilon\sigma^{\mu}\partial_{\mu}\Phi^*.
\label{super456667}
\end{eqnarray}

These supersymmetry transformations are at the origin of our approach  but the similarities stop here. In what follows based on a Lagrangian for a massive Dirac fermion with Yukawa couplings
we sketch a construction that leads to the cancellation of the quadratic divergences to the Higgs boson mass without introducing any other scalar divergences. The solution we propose can be extended to any of the particles of the standard model with the same effect. However the partners that we introduce have unusual interactions that allude them for most of the experimental search.
The final Lagrangian we obtain is symmetric in the Fourier space under a transformation that interchanges the fermions with the scalars.  This transformation is very different than the standard supersymmetric transformation and should be rather regarded as a change of variables in the path integral formalism that transforms a fermion into a boson. In a way the whole method consists in subtracting from the fermion Lagrangian the same Lagrangian expressed in terms of scalar variables.

\section{The set-up}
Consider that part of the standard model Lagrangian that contains the kinetic term for  a Dirac fermion ( for example the top quark) together with its Yukawa interaction:
\begin{eqnarray}
{\cal L}(x)=i\bar{\Psi}(x)(\gamma^{\mu}\partial_{\mu}-m)\Psi(x)-\frac{m}{v}h(x)\bar{\Psi}(x)\Psi(x),
\label{lagr546}
\end{eqnarray}
where $m$ is the mass of the fermion, $h$ is the Higgs boson and $v$ is its vacuum expectation value.

We write the corresponding action in the Fourier space:
\begin{eqnarray}
&&S=\int \frac{d^4p}{(2\pi)^4}\left[\bar{\Psi}(p)(\gamma^{\mu}p_{\mu}-m)\Psi(p)-\frac{m}{v}\int \frac{d^4q}{(2\pi)^4}h(p-q)\bar{\Psi}(p)\Psi(q)\right]=
\nonumber\\
&&=\int \frac{d^4p}{(2\pi)^4}\Bigg[\Psi_R^{\dagger}(p)\sigma^{\mu}p_{\mu}\Psi_R(p)+
\Psi_L^{\dagger}(p)\bar{\sigma}^{\mu}p_{\mu}\Psi_L(p)-m\Psi_R^{\dagger}(p)\Psi_L(p)-m\Psi_L^{\dagger}(p)\Psi_R(p)-
\nonumber\\
&&(\frac{m}{v}\int \frac{d^4q}{(2\pi)^4}h(p-q)\Psi_R^{\dagger}(p)\Psi_L(q))+h.c.)\Bigg].
\label{lagrf5467}
\end{eqnarray}

We then consider the change of variables:
\begin{eqnarray}
&&\Psi_R(p)=\sqrt{p^2-m^2}\frac{1}{\sqrt{\sigma^{\mu}p_{\mu}}}
\left[\left(
\begin{array}{c}
1\\
0
\end{array}
\right)\Phi+
\left(
\begin{array}{c}
0\\
1
\end{array}
\right)\Phi_1\right]
\nonumber\\
&&\Psi_L(p)=\sqrt{p^2-m^2}\frac{1}{\sqrt{\bar{\sigma}^{\mu}p_{\mu}}}
\left[\left(
\begin{array}{c}
1\\
0
\end{array}
\right)\varphi+
\left(
\begin{array}{c}
0\\
1
\end{array}
\right)\Phi_2\right],
\label{transf45678}
\end{eqnarray}
together with the subsequent hermitian conjugate. Here $\Phi$, $\Phi_1$, $\varphi$ and $\Phi_2$ are complex scalars that match exactly the off-shell degrees of freedom of the fermion fields. This change of variables provides a dual description of the Lagrangian in terms of new scalar variables and it is  a reminder of the process of "bozonization" (introduced by Coleman \cite{Coleman} and Mandelstam \cite{Mandelstam} for a sine-Gordon model) which up to now was known to exist for particular models but not for  a general Lagrangian.

 The above change of variables can be written in a more compact form as:
\begin{eqnarray}
&&\Psi_R=\frac{1}{xx^{\dagger}+zz^{\dagger}}(x\Phi+z\Phi_1)
\nonumber\\
&&\Psi_L=\frac{1}{yy^{\dagger}+uu^{\dagger}}(y\varphi+u\Phi_2),
\label{tarnsf44}
\end{eqnarray}
where  $x$, $z$, $y$, $u$ are:
\begin{eqnarray}
&&x=\frac{1}{\sqrt{p^2-m^2}}\sqrt{\sigma^{\mu}p_{\mu}}
\left(
\begin{array}{c}
1\\
0
\end{array}
\right)
\nonumber\\
&&z=\frac{1}{\sqrt{p^2-m^2}}\sqrt{\sigma^{\mu}p_{\mu}}
\left(
\begin{array}{c}
0\\
1
\end{array}
\right)
\nonumber\\
&&y=
\frac{1}{\sqrt{p^2-m^2}}\sqrt{\bar{\sigma}^{\mu}p_{\mu}}
\left(
\begin{array}{c}
1\\
0
\end{array}
\right)
\nonumber\\
&&u=
\frac{1}{\sqrt{p^2-m^2}}\sqrt{\bar{\sigma}^{\mu}p_{\mu}}
\left(
\begin{array}{c}
0\\
1
\end{array}
\right).
\label{rez32455}
\end{eqnarray}

Note that  the momenta depending spinors $x$, $z$, $y$ and $v$ can be taken as both commuting or anticommuting.   Most of the time we will indicate with subscripts the results for both  cases.

Below is the expression in the Fourier space of the Lagrangian in Eq. (\ref{lagr546}) in terms of the scalar variables:
\begin{eqnarray}
&&\Psi_R^{\dagger}(p)\sigma^{\mu}p_{\mu}\Psi_R(p)=
\nonumber\\
&&(x^{\dagger}\Phi^*+z^{\dagger}\Phi_1^*)\frac{1}{xx^{\dagger}+zz^{\dagger}}\sigma^{\mu}p_{\mu}\frac{1}{xx^{\dagger}+zz^{\dagger}}(x\Phi+z\Phi_1)=
\nonumber\\
&&(+_c,-_a)(p^2-m^2)\left[\Phi^*(p)\Phi(p)+\Phi_1^*(p)\Phi_1(p)\right],
\label{rez221}
\end{eqnarray}
where the subscripts $c$, $a$ refer to commuting, respectively anticommuting variables and we indicate in the round brackets the corresponding signs.
Here we used the fact that for example:
\begin{eqnarray}
&&{\rm Tr}\left[x^{\dagger}\frac{1}{xx^{\dagger}+zz^{\dagger}}]\sigma^{\mu}p_{\mu}\frac{1}{xx^{\dagger}+zz^{\dagger}}x\right]=
\nonumber\\
&&(+_c,-_a){\rm Tr}\left[\frac{1}{xx^{\dagger}+zz^{\dagger}}\sigma^{\mu}p_{\mu}\frac{1}{xx^{\dagger}+zz^{\dagger}}xx^{\dagger}\right]=p^2-m^2
\nonumber\\
&&{\rm Tr}\left[x^{\dagger}\frac{1}{xx^{\dagger}+zz^{\dagger}}\sigma^{\mu}p_{\mu}\frac{1}{xx^{\dagger}+zz^{\dagger}}z\right]=0
\label{exp76890}
\end{eqnarray}

Similarly we get:
\begin{eqnarray}
\bar{\Psi}_L(p)\bar{\sigma}^{\mu}p_{\mu}\Psi_L(p)=(+_c,-_a)(p^2-m^2)\left[\varphi^*(p)\varphi(p)+\Phi_2^*(p)\Phi_2(p)\right].
\label{rez21321}
\end{eqnarray}

Next we shall find an expression for the Yukawa term and from that we shall deduce also the mass term. We start from:
\begin{eqnarray}
&&-\frac{m}{v}h(p-q)\Psi_R^{\dagger}(p)\Psi_L(q)+h.c.=
\nonumber\\
&&-\frac{m}{v}h(p-q)\Big[(x^{\dagger}\Phi^*+z^{\dagger}\Phi_1^*)\frac{1}{xx^{\dagger}+zz^{\dagger}}|_p\frac{1}{yy^{\dagger}+uu^{\dagger}}(y\varphi+u\Phi_2)|_q\Big]+h.c.=
\nonumber\\
&&-\frac{m}{v}h(p-q)\Big[a_{p,q}\Phi^*(p)\varphi(q)+b_{p,q}\Phi^*(p)\Phi_2(q)+c_{p,q}\Phi_1^*(p)\varphi(q)+d_{p,q}\Phi_1^*\Phi_2(q)\Big]+h.c.
\label{form66789}
\end{eqnarray}
Here the subscripts denote the momenta dependence and the coefficients $a$, $b$, $c$ and $d$ are:
\begin{eqnarray}
&&a_{p,q}={\rm Tr}\left[x^{\dagger}\frac{1}{xx^{\dagger}+zz^{\dagger}}|_p\frac{1}{yy^{\dagger}+uu^{\dagger}}y|_q\right]=
\nonumber\\
&&=(+_c,-_a)\frac{\sqrt{p^2-m^2}}{p}\frac{\sqrt{q^2-m^2}}{q}{\rm Tr}\Bigg[\sqrt{\sigma^{\mu}p_{\mu}\bar{\sigma}^{\nu}q_{\nu}}\times
\left(
\begin{array}{cc}
1&0\\
0&0
\end{array}
\right)\Bigg]
\nonumber\\
&&b_{p,q}={\rm Tr}\left[x^{\dagger}\frac{1}{xx^{\dagger}+zz^{\dagger}}|_p\frac{1}{yy^{\dagger}+uu^{\dagger}}u|_q\right]=
\nonumber\\
&&=(+_c,-_a)\frac{\sqrt{p^2-m^2}}{p}\frac{\sqrt{q^2-m^2}}{q}{\rm Tr}\Bigg[\sqrt{\sigma^{\mu}p_{\mu}\bar{\sigma}^{\nu}q_{\nu}}\times
\left(
\begin{array}{cc}
0&1\\
0&0
\end{array}
\right)\Bigg]
\nonumber\\
&&c_{p,q}={\rm Tr}\left[z^{\dagger}\frac{1}{xx^{\dagger}+zz^{\dagger}}|_p\frac{1}{yy^{\dagger}+uu^{\dagger}}y|_q\right]=
\nonumber\\
&&=(+_c,-_a)\frac{\sqrt{p^2-m^2}}{p}\frac{\sqrt{q^2-m^2}}{q}{\rm Tr}\Bigg[\sqrt{\sigma^{\mu}p_{\mu}\bar{\sigma}^{\nu}q_{\nu}}\times
\left(
\begin{array}{cc}
0&0\\
1&0
\end{array}
\right)\Bigg]
\nonumber\\
&&d_{p,q}={\rm Tr}\left[z^{\dagger}\frac{1}{xx^{\dagger}+zz^{\dagger}}|_p\frac{1}{yy^{\dagger}+uu^{\dagger}}u|_q\right]=
\nonumber\\
&&=(+_c,-_a)\frac{\sqrt{p^2-m^2}}{p}\frac{\sqrt{q^2-m^2}}{q}{\rm Tr}\Bigg[\sqrt{\sigma^{\mu}p_{\mu}\bar{\sigma}^{\nu}q_{\nu}}\times
\left(
\begin{array}{cc}
0&0\\
0&1
\end{array}
\right)\Bigg],
\label{rez546789}
\end{eqnarray}
where $p=\sqrt{p^2}$ and $q=\sqrt{q^2}$.
From Eqs.  (\ref{form66789}) and (\ref{rez546789}) the counterpart for the fermion mass term emerges:
\begin{eqnarray}
&&{\cal L}_m=-m\Psi_R^{\dagger}(p)\Psi_L(p)-m\Psi_L^{\dagger}(p)\Psi_R(p)=
\nonumber\\
&&(+_c,-_a)(p^2-m^2)(-\frac{m}{p})\Big[\Phi^*(p)\varphi(p)+\varphi^*(p)\Phi(p)+\Phi_1^*(p)\Phi_2(p)+\Phi_2^*(p)\Phi_2(p)\Big],
\label{mass4344}
\end{eqnarray}
where we employed the known relations:
\begin{eqnarray}
(\sigma^{\mu}p_{\mu})(\bar{\sigma}^{\rho}p_{\rho})=(\bar{\sigma}^{\mu}p_{\mu})(\sigma^{\rho}p_{\rho})=p^2\times 1_{2\times 2}.
\label{eq21212}
\end{eqnarray}

We make the change of variables:
\begin{eqnarray}
&&\Phi=\frac{1}{\sqrt{2}}(\Phi'-\Phi'')
\nonumber\\
&&\varphi=\frac{1}{\sqrt{2}}(\Phi'+\Phi'')
\nonumber\\
&&\Phi_1=\frac{1}{\sqrt{2}}(\Phi_1'-\Phi_1'')
\nonumber\\
&&\Phi_2=\frac{1}{\sqrt{2}}(\Phi_1'+\Phi_1''),
\label{rez32212}
\end{eqnarray}
to determine the full kinetic term in the Fourier space in terms of the new variables:
\begin{eqnarray}
&&\Psi_R^{\dagger}(p)\sigma^{\mu}p_{\mu}\Psi_R(p)+\Psi_L^{\dagger}(p)\bar{\sigma}^{\mu}p_{\mu}\Psi_L(p)-(m\Psi_R^{\dagger}(p)\Psi_L(p)+h.c)=
\nonumber\\
&&(+_c,-_a)[(p^2-m^2)(1-\frac{m}{p})\Big[\Phi^{\prime*}(p)\Phi^{\prime}(p)+\Phi_1^{\prime^*}(p)\Phi_1^{\prime}(p)\Big]+(p^2-m^2)(1+\frac{m}{p})\Big[\Phi^{\prime\prime *}(p)\Phi^{\prime\prime}(p)+\Phi_1^{\prime\prime *}(p)\Phi_1^{\prime\prime}(p)\Big].
\label{rez54678}
\end{eqnarray}

\section{Properties of the scalar part of the Lagrangian}

To resume by making the simple change of variables stated in Eqs. (\ref{transf45678}) and (\ref{tarnsf44}) we obtain the scalar Lagrangian:
\begin{eqnarray}
&&{\cal L}_s=\int \frac{d^4p}{(2\pi)^4}(+_c,-_a)\Bigg[(p^2-m^2)(\Phi^*(p)\Phi(p)+\Phi_1^*(p)\Phi_1(p)+(\varphi^*(p)\varphi(p)+\Phi_2^*(p)\Phi_2(p))+
\nonumber\\
&&(p^2-m^2)(-\frac{m}{p})(\Phi^*(p)\varphi(p)+\varphi^*(p)\Phi(p)+\Phi_1^*(p)\Phi_2(p)+\Phi_2^*(p)\Phi(p))\Bigg]+
\nonumber\\
&&-\int \frac{d^4p}{(2\pi)^4}\frac{d^4q}{(2\pi)^4}\frac{m}{v}h(p-q)\Big[a_{p,q}\Phi^*(p)\varphi(q)+b_{p,q}\Phi^*(p)\Phi_2(q)+c_{p,q}\Phi_1^*(p)\varphi(q)+d_{p,q}\Phi_1^*(p)\Phi_2(q)+h.c.
\Big],
\label{fulll76878}
\end{eqnarray}
where the coefficients $a_{p,q}$, $b_{p,q}$, $c_{p,q}$ and $d_{p,q}$ are given in Eq. (\ref{rez546789}). Except for the terms in the first line of Eq. (\ref{fulll76878}) this Lagrangian cannot be put in a Lorentz invariant manner in the coordinate space and seems badly defined. We first note that the scalar on-shell states have $p^2=m^2$ and thus from the structure of the Lagrangian this means that the scalar on shell states do not interact.  However the scalar states appear as loop contributions to the Higgs boson. We will show next that these loops are very well defined and have indeed a proper Lorentz structure.

For that let us consider the couplings of the Higgs and a sample of the one loop contribution to the two point function for $h(p-q)$. This is given in essence in the path integral formalism by:
\begin{eqnarray}
&&I=2\Big[a_{p,q}\Phi^*(p)\varphi(q)+b_{p,q}\Phi^*(p)\Phi_2(q)+c_{p,q}\Phi_1^*(p)\varphi(q)+d_{p,q}\Phi_1^*(p)\Phi_2(q)\Big]\times
\nonumber\\
&&\Big[a_{p,q}\Phi^*(p)\varphi(q)+b_{p,q}\Phi^*(p)\Phi_2(q)+c_{p,q}\Phi_1^*(p)\varphi(q)+d_{p,q}\Phi_1^*(p)\Phi_2(q)\Big]^{\dagger}=
\nonumber\\
&&\Phi(p)\Phi^*(p)\Big[a_{p,q}a_{p,q}^{\dagger}\varphi(q)\varphi(q)^*+b_{p,q}b_{p,q}^{\dagger}\Phi_2(q)\Phi_2(q)^*\Big]+
\nonumber\\
&&\Phi_1(p)\Phi_1^*(p)\Big[c_{p,q}c_{p,q}^{\dagger}\varphi(q)\varphi(q)^*+d_{p,q}d_{p,q}^{\dagger}\Phi_2(q)\Phi_2(q)^*\Big].
\label{rez546788}
\end{eqnarray}
First note that   the couple $\varphi$ and $\Phi_2$ has the same propagators and different from  the couple $\Phi$ and $\Phi_1$. Let us consider in detail the quantity:
\begin{eqnarray}
&&\langle[a_{p,q}a_{p,q}^{\dagger}\varphi(q)\varphi(q)^{\dagger}+b_{p,q}b_{p,q}^{\dagger}\Phi_2(q)\Phi_2(q)^*]\rangle=
\nonumber\\
&&\langle \varphi(q)\varphi^*(q)\rangle\Big[[x_p^{\dagger}\frac{1}{\sigma^{\mu}p_{\mu}}\frac{1}{\bar{\sigma}^{\nu}q_{\nu}}y_q y_q^{\dagger}
\frac{1}{\bar{\sigma}^{\nu}q_{\nu}}\frac{1}{\sigma^{\mu}p_{\mu}}x_p]+
[x_p^{\dagger}\frac{1}{\sigma^{\mu}p_{\mu}}\frac{1}{\bar{\sigma}^{\nu}q_{\nu}}u_q u_q^{\dagger}
\frac{1}{\bar{\sigma}^{\nu}q_{\nu}}\frac{1}{\sigma^{\mu}p_{\mu}}x_p]\Big]=
\nonumber\\
&&(+_c,-_a){\rm Tr}\Big[\langle \varphi(q)\varphi^*(q)\rangle[\frac{1}{\sigma^{\mu}p_{\mu}}\frac{1}{\bar{\sigma}^{\nu}q_{\nu}}(y_q y_q^{\dagger}+u_q u_q^{\dagger})
\frac{1}{\bar{\sigma}^{\nu}q_{\nu}}\frac{1}{\sigma^{\mu}p_{\mu}}x_px_p^{\dagger}]\Big]=
\nonumber\\
&&\langle \varphi(q)\varphi^*(q)\rangle(+_c,-_a)(q^2-m^2){\rm Tr}\Big[\frac{1}{\sigma^{\mu}p_{\mu}}\frac{1}{\bar{\sigma}^{\nu}q_{\nu}}\bar{\sigma}^{\rho}q_{\rho}
\frac{1}{\bar{\sigma}^{\nu}q_{\nu}}\frac{1}{\sigma^{\mu}p_{\mu}}x_px_p^{\dagger}\Big]=
\nonumber\\
&&\langle \varphi(q)\varphi^*(q)\rangle(+_c,-_a)(q^2-m^2)(p^2-m^2)\frac{1}{q^2p^2}{\rm Tr}\Big[\sigma^{\mu}q_{\mu}\sqrt{\bar{\sigma}^{\nu}p_{\nu}}\times
\left(
\begin{array}{cc}
1&0\\
0&0
\end{array}
\right)
\sqrt{\bar{\sigma}^{\nu}p_{\nu}}\Big].
\label{contr876889}
\end{eqnarray}
Note that here we used the fact that the behavior of $\varphi$ and $\Phi_2$ is identical such that  $\langle \varphi^*(q)\varphi \rangle=\langle \Phi_2^*(q)\Phi_2 \rangle$.
Keeping in mind that the above contribution is coming from $\Phi(p)\Phi(p)^*$ and there is a similar contribution with the same propagator coming from $\Phi_1(p)^*\Phi_1(p)$ one obtains:
\begin{eqnarray}
&&\langle I \rangle=2\langle \Phi^*(p)\Phi(p)\rangle \langle \varphi(q)\varphi^*(q)(+_c,-_a)(q^2-m^2)(p^2-m^2)\frac{1}{p^2q^2}\times
\nonumber\\
&&{\rm Tr}\Big[\sigma^{\mu}q_{\mu}\sqrt{\bar{\sigma}^{\nu}p_{\nu}}\times
\left(
\begin{array}{cc}
1&0\\
0&0
\end{array}
\right)
\sqrt{\bar{\sigma}^{\nu}p_{\nu}}\Big]+
{\rm Tr}\Big[\sigma^{\mu}q_{\mu}\sqrt{\bar{\sigma}^{\nu}p_{\nu}}\times
\left(
\begin{array}{cc}
0&0\\
0&1
\end{array}
\right)
\sqrt{\bar{\sigma}^{\nu}p_{\nu}}\Big]=
\nonumber\\
&&=\langle \Phi^*(p)\Phi(p)\rangle \langle \varphi(q)\varphi^*(q)\rangle (+_c,-_a)(q^2-m^2)(p^2-m^2)\frac{1}{q^2p^2}{\rm Tr}[\sigma^{\mu}p_{\mu}\bar{\sigma}^{\nu}q_{\nu}].
\label{rez43567}
\end{eqnarray}
Note that this loop contribution makes perfect sense from the point of view of the Lorentz properties. However this should be considered only as an exercise that shows the proper behavior of the scalars in loops since the correct two point contributions to the Higgs mass should be calculated in terms of the mass eigenstates in the Lagrangian which are the fields $\Phi'$, $\Phi''$, $\Phi_1'$ and $\Phi''$.

The full Lagrangian (wit both the fermion and the scalar terms) is invariant under the transformations in the Fourier space:
\begin{eqnarray}
&&\delta \Psi_R(p)=(+_a,-_c)\sqrt{p^2-m^2}\frac{1}{\sqrt{\sigma^{\mu}p_{\mu}}}\Big[
\left(
\begin{array}{c}
1\\
0
\end{array}
\right)\Phi(p)\epsilon+
\left(
\begin{array}{c}
0\\
1
\end{array}
\right)\Phi_1(p)\epsilon\Big]
\nonumber\\
&&\delta\Psi_L=(+_a,-_c)\sqrt{p^2-m^2}\frac{1}{\sqrt{\bar{\sigma}^{\mu}p_{\mu}}}\Big[
\left(
\begin{array}{c}
1\\
0
\end{array}
\right)\varphi(p)\epsilon+
\left(
\begin{array}{c}
0\\
1
\end{array}
\right)\Phi_2(p)\epsilon\Big]
\nonumber\\
&&\Phi(p)=(+_a,-_c)\frac{1}{\sqrt{p^2-m^2}}\sqrt{\sigma^{\mu}p_{\mu}}(1,0)\Psi_R\epsilon
\nonumber\\
&&\Phi_1(p)=(+_a,-_c)\frac{1}{\sqrt{p^2-m^2}}\sqrt{\sigma^{\mu}p_{\mu}}(0,1)\Psi_R\epsilon
\nonumber\\
&&\varphi(p)=(+_a,-_c)\frac{1}{\sqrt{p^2-m^2}}\sqrt{\bar{\sigma}^{\mu}p_{\mu}}(1,0)\Psi_L\epsilon
\nonumber\\
&&\Phi_2(p)=(+_a,-_c)\frac{1}{\sqrt{p^2-m^2}}\sqrt{\bar{\sigma}^{\mu}p_{\mu}}(0,1)\Psi_L\epsilon,
\label{tarnsf8879}
\end{eqnarray}
where $\epsilon$ is an arbitrary constant scalar. These relations should be taken together with their subsequent hermitian conjugate counterparts. This invariance can be checked directly very easily but here we will only note here that this is evident as a mere consequence of the fact that we just subtract from the fermion Lagrangian the same Lagrangian expressed in terms of the scalar variables so the transfomations in eq. (\ref{tarnsf8879}) are just a change of variables.  Although this transformations change a fermion into a boson and the reciprocal they should not be regarded as  a supersymmetry transformations.

\section{ A solution to the naturalness of the Higgs boson}

The Higgs boson $h$ of the full Lagrangian is free of quadratic divergences. This stems on general grounds from the construction of the scalar Lagrangian.
To make matter more clearly we shall prove directly so in what follows.

First we shall consider the quadratic divergences to the Higgs boson  mass stemming from the tadpole diagrams. The corresponding fermion contribution to $\Sigma(0)$ apart for some factor ahead is given by:
\begin{eqnarray}
-i\frac{m}{v}(-i){\rm Tr}\int \frac{d^4p}{(2\pi)^4}\frac{\gamma^{\mu}p_{\mu}+m}{p^2-m^2}=\frac{-m^2}{v}\int\frac{d^4p}{(2\pi)^4}\frac{4}{p^2-m^2}=\frac{4m^2}{v}\frac{i}{16\pi^2}\Lambda^2+
{\rm log\,\,terms}.
\label{quadr443}
\end{eqnarray}
Before going further we need to establish the propagators of the scalar particles in the mass eigenstate basis:
\begin{eqnarray}
&&{\rm Propagator\,}\Phi'=(-_a,+_c)\frac{i}{(p^2-m^2)(1-\frac{m}{p})}
\nonumber\\
&&{\rm Propagator\,}\Phi''=(-_a,+_c)\frac{i}{(p^2-m^2)(1+\frac{m}{p})}
\nonumber\\
&&{\rm Propagator\,}\Phi_1'=(-_a,+_c)\frac{i}{(p^2-m^2)(1-\frac{m}{p})}
\nonumber\\
&&{\rm Propagator\,}\Phi_1''=(-_a,+_c)\frac{i}{(p^2-m^2)(1+\frac{m}{p})}.
\label{prop98987}
\end{eqnarray}
Then the tadpole scalar contribution to $\Sigma(0)$ is given by:
\begin{eqnarray}
&&-i(-_a,+_c)\int \frac{d^4p}{(2\pi)^4}\frac{m}{v}\frac{p^2-m^2}{p}(-_a,+_c)\left[\frac{2i}{(p^2-m^2)(1-\frac{m}{p})}-\frac{2i}{(p^2-m^2)(1+\frac{m}{p})}\right]=
\nonumber\\
&&2\int \frac{d^4p}{(2\pi)^4}\frac{m}{v}p^2\frac{p^2-m^2}{p}\frac{(1+\frac{m}{p})-(1-\frac{m}{p})}{(p^2-m^2)^2}=
\nonumber\\
&&4\frac{m^2}{v}\int \frac{d^4p}{(2\pi)^4}\frac{1}{p^2-m^2}=-\frac{4m^2}{v}\frac{i}{16\pi^2}\Lambda^2+
{\rm log\,\,terms}.
\label{scq3232}
\end{eqnarray}

Next we consider the 1 PI contributions  to the two point function of the Higgs boson $i\Sigma(0)$. The fermion loop yields apart for some factors ahead:
\begin{eqnarray}
&&-\frac{m^2}{v^2}(-1) {\rm Tr}\int \frac{d^4p}{(2\pi)^4}\frac{i}{\gamma^{\mu}p_{\mu}-m}\frac{i}{\gamma^{\nu}p_{\nu}-m}=
\nonumber\\
&&-\frac{4m^2}{v^2}\int \frac{d^4p}{(2\pi)^4}\frac{p^2+m^2}{(p^2-m^2)^2}=\frac{4m^2}{v}\frac{i}{16\pi^2}\Lambda^2+
{\rm log\,\,terms}.
\label{rez3232}
\end{eqnarray}
 The scalar contribution is given by:
\begin{eqnarray}
&&-\int \frac{d^4p}{(2\pi)^4}\frac{m^2(p^2-m^2)^2}{p^2}\left[\frac{-2}{(p^2-m^2)^2(1-\frac{m}{p})^2}+\frac{-2}{(p^2-m^2)^2(1+\frac{m}{p})^2}\right]=
\nonumber\\
&&\frac{2m^2}{v^2}\int \frac{d^4p}{(2\pi)^4}\frac{(p^2-m^2)^2}{p^2}p^4\frac{(1-\frac{m}{p})^2+(1+\frac{m}{p})^2}{(p^2-m^2)^4}=
\nonumber\\
&&\frac{2m^2}{v^2}\int \frac{d^4p}{(2\pi)^4}\frac{(p-m)^2+(p+m)^2}{(p^2-m^2)^2}=\frac{4m^2}{v^2}\int \frac{d^4p}{(2\pi)^4}\frac{p^2+m^2}{(p^2-m^2)^2}
=-i\frac{4m^2}{v^2}\frac{1}{16\pi^2}\Lambda^2+
{\rm log\,\,terms}.
\label{rez3232}
\end{eqnarray}

As expected the one loop corrections to the Higgs two point function cancel each other exactly (even at the level of logarithms) so the Lagrangian is free of quadratic divergences to the Higgs boson mass.

\section{Conclusions}

The main point of the present work is to propose a solution to the naturalness problem that does not suffer from any of experimental setbacks of the alternative theories. This is based on a change of variables in the Fourier space that transforms a fermion into a boson and viceversa. Although inspired by supersymmetry this change of variables is not associated with any symmetry of the space time. The full Lagrangian containing both the initial particles and the transformed ones is free of quadratic divergences. We showed this at one loop but the intrinsic structure of the Lagrangian guarantees this at all orders.

The partners of the standard particles participate fully as quantum fluctuations in any quantum corrections and interactions. But by construction of the Lagrangian the partners do not appear as on-shell states (final states) for any of the interactions with the standard particles. There is a possibility that this constraint will be shifted for some interactions with other partners.

We illustrate our approach based on the Lagrangian of a massive Dirac fermion coupled with a Higgs through Yukawa interactions. The method can be extended in principle to any particle and interaction of the standard model. In this case possible dark matter candidates might emerge of some of the partners. Our purpose here was just to introduce a solution to the electroweak
naturalness problem and show how this works in detail. A detailed analysis of the full standard model form this point of view and also a complete study of the phenomenological consequences will be made in a future work.

\section*{Acknowledgments} \vskip -.5cm

The work of R. J. was supported by a grant of the Ministry of National Education, CNCS-UEFISCDI, project number PN-II-ID-PCE-2012-4-0078.

\end{document}